\documentclass[fleqn,usenatbib]{mnras}

\usepackage{mathptmx}
\usepackage{color}
\usepackage[T1]{fontenc}

\usepackage{booktabs}

\DeclareRobustCommand{\VAN}[3]{#2}
\let\VANthebibliography\thebibliography
\def\thebibliography{\DeclareRobustCommand{\VAN}[3]{##3}\VANthebibliography}

\usepackage{graphicx}	% Including figure files
\usepackage{amsmath}	% Advanced maths commands
\usepackage{amssymb}	% Extra maths symbols
\usepackage{multirow}
\usepackage{rotating}

\usepackage{caption}
\usepackage{subcaption}

\newcommand{\msun}{M$_\odot$}
\defcitealias{Gruyters2014}{G14}
\newcommand{\teff}{T$_{\mathrm{eff}}$}

\title[Lithium abundances in the GC NGC~6752]{Lithium abundances as a tracer of AGB stars pollution in the globular cluster NGC~6752}

\author[J. Schiappacasse-Ulloa et al.]{
Schiappacasse-Ulloa, J.,$^{1,}$$^{2}$\thanks{E-mail: joseluis.schiappacasseulloa@studenti.unipd.it}
Sara Lucatello,$^{2,3}$
Rain, M. J.,$^{1}$
Adriano Pietrinferni$^{4}$
\\
$^{1}$Dipartimento di Fisica e Astronomia, Universita’ di Padova, Vicolo dell' Osservatorio 3, I-35122, Padova, Italy\\
$^{2}$INAF–Osservatorio Astronomico di Padova, Vicolo dell’Osservatorio 5, 35122 Padova, Italy\\
$^{3}$Institute for Advanced Studies, Technische Universit{\"a}t M{\"u}nchen, Lichtenbergstra{\ss}e 2 a, 85748 Garching bei M{\"u}nchen\\
$^{4}$INAF–Osservatorio Astronomico d’Abruzzo, Via M. Maggini, s/n, I-64100, Teramo, Italy \\
}

\date{Accepted XXX. Received YYY; in original form ZZZ}

\pubyear{2021}

\begin{document}
\label{firstpage}
\pagerange{\pageref{firstpage}--\pageref{lastpage}}
\maketitle

\begin{abstract}
This paper presents the chemical abundance analysis of 217 stars in the metal-poor globular cluster NGC 6752, distributed from the turn-off to the lower red giant branch. Aluminium and Lithium abundances were derived through spectral synthesis applied to spectra collected with FLAMES, both in GIRAFFE and UVES modes. 
The work aims to gain insight into the nature of the polluter(s) responsible for the abundance variations and C-N, Na-O, Al-Mg anti-correlations associated with the multiple population phenomenon. 
We found a plateau at A(LI) $=$2.33$\pm$0.06 dex in unevolved stars, with the average Li content decreasing continuously down to $\sim$1.25 dex at the bottom of the red giant branch. As expected in the classic anti-correlation scenario, we found stars low in Al and high Li abundance and stars high in Al and low in Li.
However, in addition, we also found evidence of Al-rich, second-generation stars with high Li content. This finding suggests the need for Li production, known to happen in intermediate-mass ($\sim$4-8\,\msun) AGB stars through the Cameron-Fowler mechanism.
It is worth noticing that the Li abundance observed in Al-rich stars never exceeds that in Al-poor stars.
\end{abstract}


%%%%%%%%%%%%%%%%%%%%%%%%%%%%%%%%%%%%%%%%%%%%%%%%%%

%%%%%%%%%%%%%%%%% BODY OF PAPER %%%%%%%%%%%%%%%%%%

\section{Introduction}
\label{Sec:Intro}

%\textcolor{magenta}{This part is not particularly relevant here}
Globular clusters (GCs) are fundamental laboratories to study stellar evolution and gain insight into old stellar populations. Their study is also crucial in the understanding of the formation of the Milky Way. Their contribution to the formation of the Halo is considerable and possibly also for the  Bulge  \citep{Martell_2011,Schiavon_2020}.
Therefore, the understanding of the processes that lead to their formation and the so-called multiple stellar populations (MSP) phenomenon, is relevant in the broad context of probing the formation of the Galactic components. 

The MSP phenomenon has been detected through both photometric \citep[see e.g.,][]{Monelli2013} and spectroscopic \citep[see e.g.,][]{Carretta2009b} studies. Parallel evolutionary sequences can be seen in their color-magnitude diagrams (CMDs) in the former. In the latter, chemical abundance patterns of light elements  (e.g., C, N, O, Na, Mg, Al) show variations in their different populations.

The abundance of these elements are characterised by anti-correlations, C anti-correlates with N, Na with O, and Mg with Al \citep[see e.g.,][]{Carretta2009}, that extend with the same range all the way to the main sequence \citep{Gratton_2001} and are hence present since birth, thus reflecting the composition of the gas they formed from, suggesting a scenario of self-pollution.

As was explored in several literature studies (e.g., \citealt{Arnould1999, Charbonnel2016}), these patterns can be the product of nucleosynthesis happening in different sites. In particular, the so-called hot H-burning can produce N, Na, and Al through the  CN, NeNa, and MgAl cycles. Each of them requires minimum activation T$_{\mathrm{eff}}$'s of 10~MK, 40~MK, and 70~MK, respectively.
 
Currently, the most accepted scenario describes a group of stars, the so-called First Generation (FG) stars, polluting pristine gas (i.e., the same gas from which the FG formed). A new generation of stars is formed from the mixed material with an altered composition in the mentioned elements \citep{Bastian2018, Gratton2019}, the so-called second-generation stars (SG).

The group(s) of stars responsible for the pollution is(are) still under debate and several polluters have been proposed as the source of the internal light-element variations found in GCs, such as fast-rotating massive stars (FRMS; \citealt{Meynet2006}), massive binaries \citep{DeMink2009}, supermassive stars (M $> 10^4$\,\msun; \citealt{Denissenkov_2015}), and intermediate-mass ($\sim$4-8\,\msun) asymptotic giant branch (AGB; \citealt{Ventura2001}) stars. 

In this context, Li can provide helpful information to probe this issue. Its low burning temperature ($\sim2.5\times10^{6}$~K) means that Li is destroyed whenever the NeNa (and even more so the MgAl) cycle is activated. Therefore, the polluting material processed through hot H-burning, rich in N, Na, and Al and depleted in C, O, and Mg, should have negligible Li content, and the SG stars formed from gas enriched with this material should have low Li since birth.

While a large number of clusters have by now been studied their p-capture elements abundances insofar, only a few studies have measured Li in conjunction with other light elements involved in (anti)correlations. This is due to the fact that Li is undetectable in stars brighter than the bump. Therefore, the study of Li requires spectra of stars considerably fainter than the bright giants generally used in the study of the chemistry of GCs, and in turn a proportionally considerable increase in observing time. 
NGC~6121 \citep{Dorazi2010,Mucciarelli20111}, NGC~6218 \citep{dorazi2014}, and NGC~362 \citep{Dorazi2015} have shown that Li-rich stars were unexpectedly present among their SG population. 
\citet{Shen2010} analysed a large sample of turn-off (TO) stars in NGC~6752. They found a Li-O correlation with a shallower slope than expected, which they interpret as a need for Li production in the polluters.

Later, \citet[hereafter G14]{Gruyters2014} confirmed the Li-Na anti-correlation in both evolved and unevolved members of the same cluster. They also reported a fraction of intermediate population (mildly Al-rich) stars with Li abundance similar to the abundance of FG stars.

The anti-correlations in G14 is what is expected in a simple self polluting scenario, as material processed by one (or more) of the polluters (e.g., intermediate and massive 3-5\,\msun AGB stars, FRMS, supermassive stars, and massive binaries) would enrich the environment with hot H-burning processed material: rich in N, Na and Al and poor in C, O and Mg and utterly devoid of Li.
On the other hand, detecting Li-rich stars among Al-rich stars (or Na-rich) would suggest the need for Li production, hence the contribution of a mechanism/nucleosynthetic site capable of Li production. Given that among the candidate polluters responsible for the MSP phenomenon, only AGB stars are known to be capable of non-negligible Li production, such finding would indicate that AGB stars were (one of) the polluters. 
In this context, NGC~6752 is of particular interest. It is one of the clusters known to host a population with a considerable spread in lithium, also among SG stars, which is of particular interest \citep{Shen2010}. NGC~6752 is an ideal candidate to investigate potential systematic differences between both Li-poor and Li-rich stars. We expect to characterise the different stellar populations of the cluster in terms of Li.

While previous studies for this cluster are available in the literature (see discussion above), we aim to analyse all the available archival data homogeneously, from the turn-off to the RGB \textit{bump}, to paint a comprehensive picture of Li and Al in the cluster.

In \S\ref{Sec:Observation}, we describe the target selection and observation. In \S\ref{Sec:Data Analysis}, we described the stellar parameter's determination and data analysis. We present our results and discussion in \S\ref{Sec:Results}. Finally, in \S\ref{Sec:Conclusion}, we look into the conclusions.

\section{Target selection and Observation}
\label{Sec:Observation}
We analysed the spectra of 217 stars of NGC~6752 among TO, sub-giant (SGB), and red giant branch (RGB) stars. The spectra of 126  stars, previously analysed by \citetalias{Gruyters2014}, were kindly provided to us by the authors. They used spectra observed with the mid-resolution spectrograph FLAMES/GIRAFFE using the setup HR15N (6470-6790~\r{A}, $R\equiv \lambda/\Delta\lambda$=~19200) under the ESO-VLT  programs 079.D-0645(A) and 081.D-0253(A).

The spectra of the remaining 91 stars were downloaded from the Gaia-ESO collection of the ESO archive. They are a combination of FLAMES/GIRAFFE (HR15N: 6444-6816~\r{A} with $R \equiv$~19200) and FLAMES/UVES spectra (4768-5801~\r{A} and 5822-6830~\r{A} with $R \equiv$ 47000 each). The cluster members were selected based on  \emph{Gaia}~eDR3 \citep{Gaia_2020} data. Only stars within 3$\sigma$ on parallax and proper motions values from the cluster mean value were considered members.
The wavelength range covered by the presently analysed GIRAFFE and UVES spectra allows measuring both Li doublet at 6708 \r{A} and Al lines at 6696 \r{A} and 6698 \r{A}.

\begin{table*}
   \centering
   \caption{Stellar parameters, [Fe/H] (adopted from \citet{Carretta2009c}), A(Li), and [Al/Fe] for TO, SGB, and RGB stars observed in NGC 6752. This table is fully available in electronic format. A(Al)$_{\odot}$=6.41}
   \label{tab:summary_1}
\begin{tabular}{ccccccccccccc}\\
       \hline
       \hline
    ID   & T$_{eff}$  & $\sigma$ & $\log g$  & $\sigma$ & [Fe/H] & $\sigma$  & v$_{t}$ & $\sigma$ & A(Li) & $\sigma$ & [Al/Fe] & $\sigma$ \\
       \hline
2445 & 5594 & 40 & 3.66 & 0.05 & -1.56 & 0.01 & 1.04 & 0.02 & 1.20 & 0.06 & 0.58  & 0.10\\
2472 & 5513 & 12 & 3.55 & 0.00 & -1.56 & 0.01 & 1.08 & 0.00 & 1.48 & 0.05 & 0.22  & 0.10\\
2476 & 5586 & 15 & 3.62 & 0.00 & -1.56 & 0.01 & 1.05 & 0.00 & 1.42 & 0.05 & 0.54  & 0.10\\
2497 & 5545 & 21 & 3.61 & 0.06 & -1.56 & 0.01 & 1.06 & 0.02 & 1.78 & 0.05 & -0.16 & 0.10\\
2517 & 5567 & 20 & 3.62 & 0.01 & -1.56 & 0.01 & 1.05 & 0.00 & 1.65 & 0.05 & -0.68 & 0.10\\
...  & ...  & ...& ...  & ...  &  ...  & ...  & ...  & ...  & ...  & ...  & ...   & ... \\
       \hline
\end{tabular}
\end{table*}

\begin{table}
   \centering
   \caption{The number of stars for which Li and Al are measured among TO/SGB and RGB stars. The number in parentheses indicates upper limits. The last column indicates the total number of objects for which spectra were analysed.}
   \label{tab:summary_2}
\begin{tabular}{lrrrrr}\\
       \hline
       \hline
       & & Instrument  & Li       & Al       & number   \\
       \hline
\multirow{2}{2em}{\rotatebox{0}{TO/SGB}}    & & GIRAFFE  & 115 (37)  & 12 (140)  & 152 \\
       & & UVES  & 4 (0) & 0 (4) & 4 \\
       \multirow{2}{2em}{\rotatebox{0}{RGB}} & & GIRAFFE  & 28 (26) & 20 (34) & 54 \\
       & & UVES  & 6 (1) & 2 (5) & 7 \\
       \hline
\end{tabular}
\end{table}

\begin{figure}
	\includegraphics[width=\columnwidth]{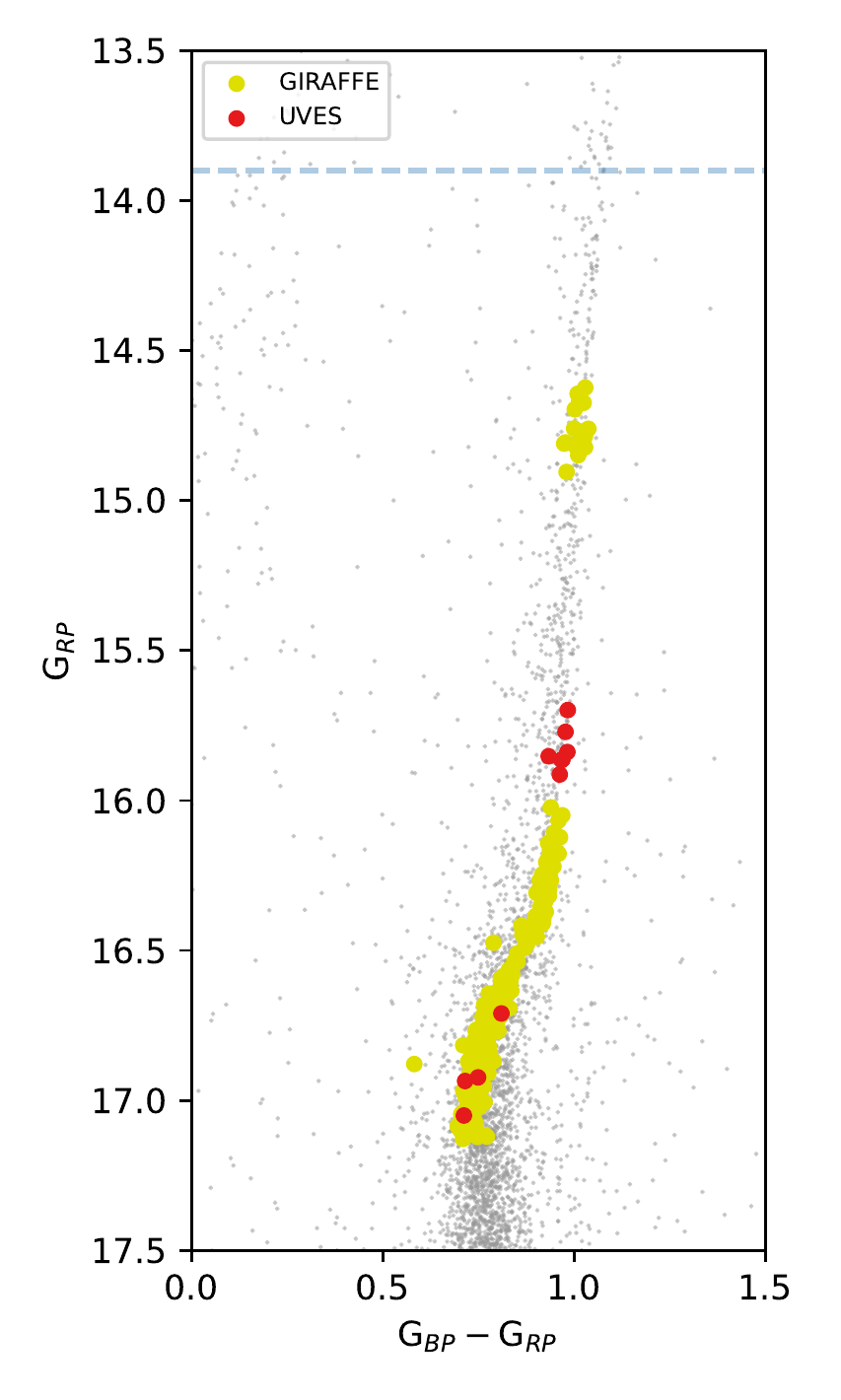}
    \caption{CMD of NGC 6752. Red and yellow dots represent UVES and GIRAFFE spectra, respectively. Gray dots are photometric data from Gaia early DR3 in the cluster field. The blue horizontal dashed line shows the RGB-bump.}
    \label{fig:cmd}
\end{figure}

Figure \ref{fig:cmd} shows the final sample in the CMD. GIRAFFE and UVES spectra are represented with yellow and red dots, respectively, all of them are located between the turn-off point (TOP) and the RGB-bump (blue dashed line), where we expect non-negligible Li abundances. Li is highly depleted above the bump and not detectable in the corresponding spectra. Hence we have not included stars of this evolutionary stage in our sample.

We used \texttt{IRAF} \footnote{\texttt{IRAF} is the Image Reduction and Analysis Facility, a general purpose software system for the reduction and analysis of astronomical data. \texttt{IRAF} is written and supported by National Optical Astronomy Observatories (NOAO) in Tucson, Arizona.} to perform the continuum normalization,  measure the radial velocity (V$_{r}$), and shift to rest-frame the spectra downloaded from the ESO archive. The \citetalias{Gruyters2014} spectra were provided at rest-frame and continuum normalised.
We found a $<$V$_{r}$ $>$= -26.14$\pm$0.50 km s$^{-1}$, based on 91 individual stars. Our radial velocity is in excellent agreement with the one reported in the Harris Catalog\footnote{https://www.physics.mcmaster.ca/~harris/mwgc.dat} \cite{Harris1996}, which is V$_{r}$ = -26.70$\pm$0.75 km s$^{-1}$. 

\section{Data Analysis}
\label{Sec:Data Analysis}

Our sample includes objects covering an extensive range of evolutionary stages, and the derivation of the homogeneous atmospheric parameters is crucial in the analysis.

\subsection{Photometric Stellar Parameters}

We derived the stellar parameters from photometry as follows. Color-temperature relations are generally provided separately for unevolved and evolved stars. This is the case of two commonly used temperature scales, reported by \citet{Alonso1996} and \citet{Alonso1999}. In order to make sure that photometric effective temperatures (T$_{\mathrm{eff}}$) are on a homogeneous scale for both TO/SGB and RGB stars, we adopted the corrections described in \citet{Korn2007}, based on the \texttt{$(v - y)$} color from Str\"{o}mgren photometry \citet{Grundahl1999} using E(B-V) = 0.04~mag \citep{Harris1996}. The reddening has been transformed in the appropriate bands following \citet{Crawford1975}. To find the $\log g$  --- corresponding to the photometric temperatures of our targets --- we used the isochrone from \citet{Bressan2012}, adopting an age of 13.5~Gyr \citep{Gruyters2013}, and [Fe/H]=-1.56 dex \citep{Carretta2009c}. Finally, we used the formula\footnote{v$_{t} = 2.22 - 0.322$ $\log g$} described in \citet{Gratton1999} to estimate the microturbulence velocity (v$_{t}$) of the stars.

\subsection{Spectroscopic Stellar Parameters}

\begin{figure}
	\includegraphics[width=\columnwidth]{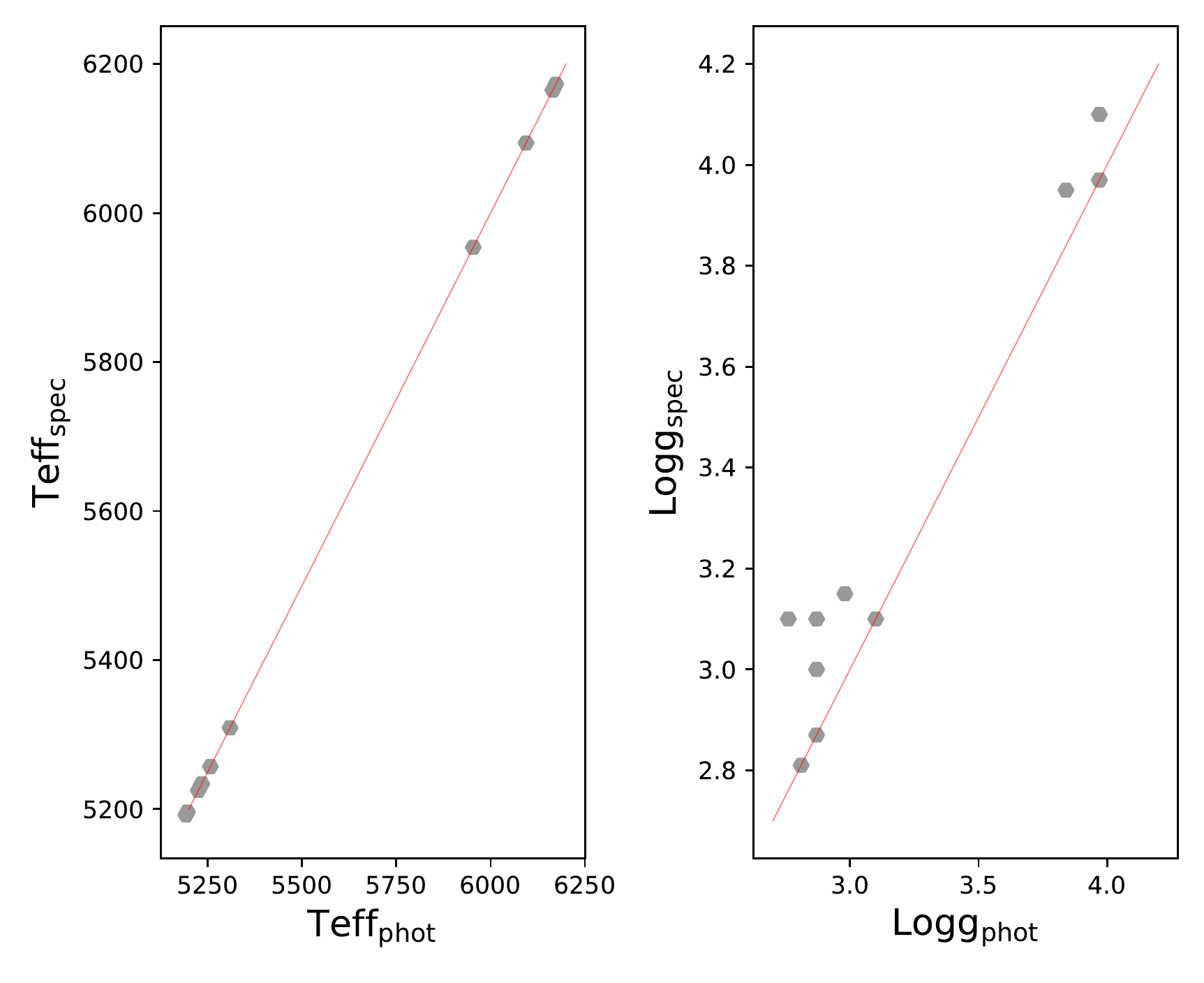}
    \caption{Comparison between the stellar parameters obtained from photometry and spectroscopy.}
    \label{fig:spec_phot}
\end{figure}

The GIRAFFE spectra for our sample have somewhat limited coverage in wavelength \footnote{A fraction of the stars have two orders available in the archive and an even smaller fraction has more, but we are only considering HR15 here as we are interested in performing a fully consistent analysis}, and hence a limited number of Fe lines, leading to considerable uncertainties in the determination of atmospheric parameters from spectroscopy. On the other hand, the high-quality Str\"{o}mgren photometry from one source provides precise colours and magnitudes, thus minimising random errors. Hence our approach is to adopt the photometric T$_{\mathrm{eff}}$, $\log g$, and v$_{t}$.

For UVES spectra, which have a much broader wavelength coverage, stellar parameters, can on the other hand, be reliably derived spectroscopically, using the photometric ones as an initial guess. We performed a traditional spectroscopic analysis to derive the atmospheric parameters. Equivalent widths for iron lines were measured using \texttt{ARES} \citep{Sousa2007}, which were analysed later using \texttt{abfind} driver from \texttt{MOOG}\footnote{We used \texttt{pyMOOGi} version November 2019. It can be downloaded from: \url{https://github.com/madamow/pymoogi}} \citep{Sneden1973}, a 1-D LTE line analysis code, to get the final set of stellar parameters, and the \citet{Kurucz1992} grids of model atmosphere. The comparison between photometric and spectroscopic stellar parameters is shown in Figure \ref{fig:spec_phot}. As can be seen, T$_{\mathrm{eff}}$ obtained using both methods are in overall excellent agreement. Spectroscopic gravities are on average 0.10$\pm$0.01 dex larger than their photometric counterparts.
Nevertheless, we adopted the photometric set of stellar parameters even for UVES stars to have complete consistency across the whole sample and minimise internal errors.

\begin{figure}
	\includegraphics[width=\columnwidth]{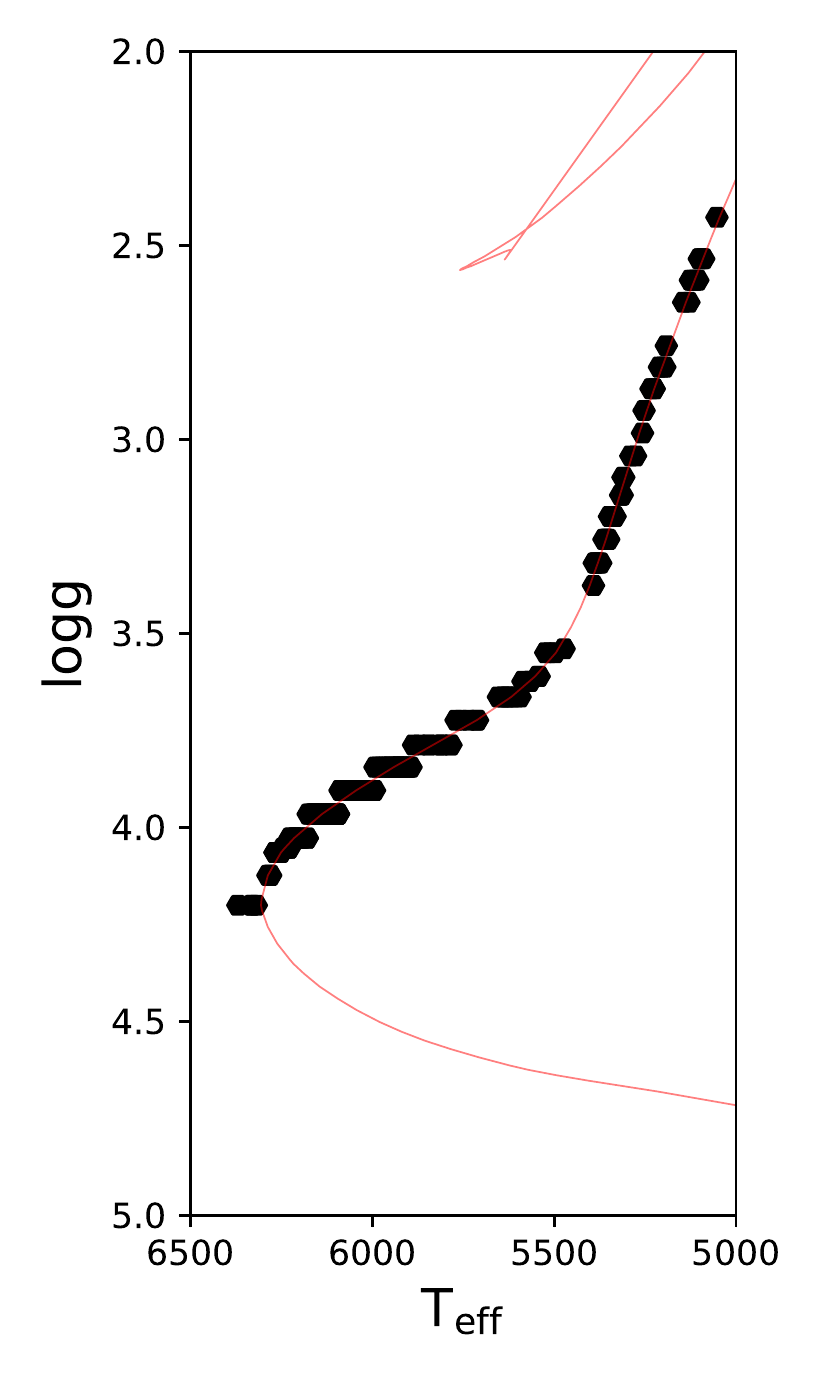}
    \caption{Distribution of our sample in the T$_{\mathrm{eff}}$ -- $\log$ g plane. Black filled hexagons represent our T$_{\mathrm{eff}}$ and $\log$g. The red line shows their theoretical values from isochrones.}
    \label{fig:theo_stellar_parameters}
\end{figure}

Figure \ref{fig:theo_stellar_parameters} shows the adopted stellar parameters for the whole sample in the T$_{\mathrm{eff}}$ -- $\log$g plane. The black filled hexagons represent our targets and the red line is the theoretical T$_{\mathrm{eff}}$, and $\log$g from isochrones.

\subsection{T$_{\mathrm{eff}}$~--~$\log$g Comparison with \citetalias{Gruyters2014}}

\begin{figure}
	\includegraphics[width=\columnwidth]{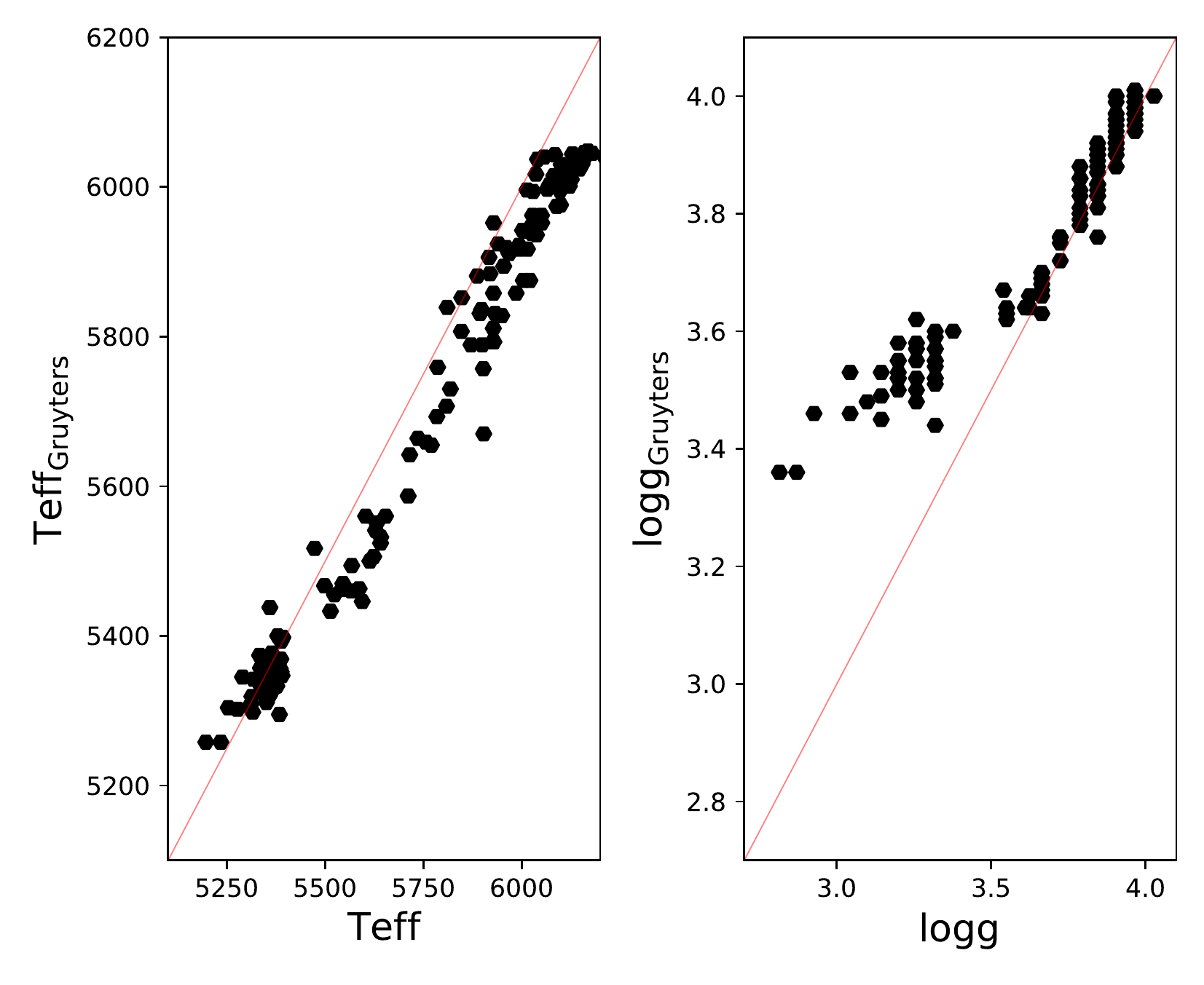}
    \caption{Comparison between our T$_{\mathrm{eff}}$ and $\log$g and the ones reported by \citetalias{Gruyters2014}. Red line indicate a one-to-one relation.}
    \label{fig:comparison_ste_par}
\end{figure}

As discussed above, 126 of the stars in our sample are in common with \citetalias{Gruyters2014} (they present the analysis for 193 NGC~6752 stars, but only 126 have the spectral coverage suitable for our purposes). In Figure \ref{fig:comparison_ste_par}, we compare our stellar parameters with those reported by them. They used the \citep{Korn2007} modified color--T$_{\mathrm{eff}}$ relations of \citet{Alonso1996} and \citet{Alonso1999} for both TO and RGB stars. The temperatures were linearly interpolated for SGB. While there is a good agreement in T$_{\mathrm{eff}}$ for evolved stars, we systematically reported lower T$_{\mathrm{eff}}$ ($\sim$ 80 K) in the unevolved ones. The reason is likely due to their infrared flux method calibration to the derivated T$_{\mathrm{eff}}$ and slightly lower adopted ([Fe/H]=-1.60 dex) metallicity.
We also have different approaches in the $\log$g determination. While we obtained it directly from isochrones, they use canonical formula (e.g., equation 1 in \citealt{Rain2019}). As can be seen from the Figure, we have good agreement among unevolved stars, but a considerable difference between our $\log$g among the evolved ones (mean difference of 0.3 dex). We note, however, that the species in which we are interested are not strongly dependent on these parameters (see table \ref{tab:sensitivity_matrix}). It is important to note, however, that for the purpose of this paper, the priority is to be fully homogeneous and internally consistent, and that consistency with the literature is of lesser importance.

\begin{figure*}
     \centering
     \begin{subfigure}[b]{0.45\textwidth}
         \centering
         \includegraphics[width=\textwidth]{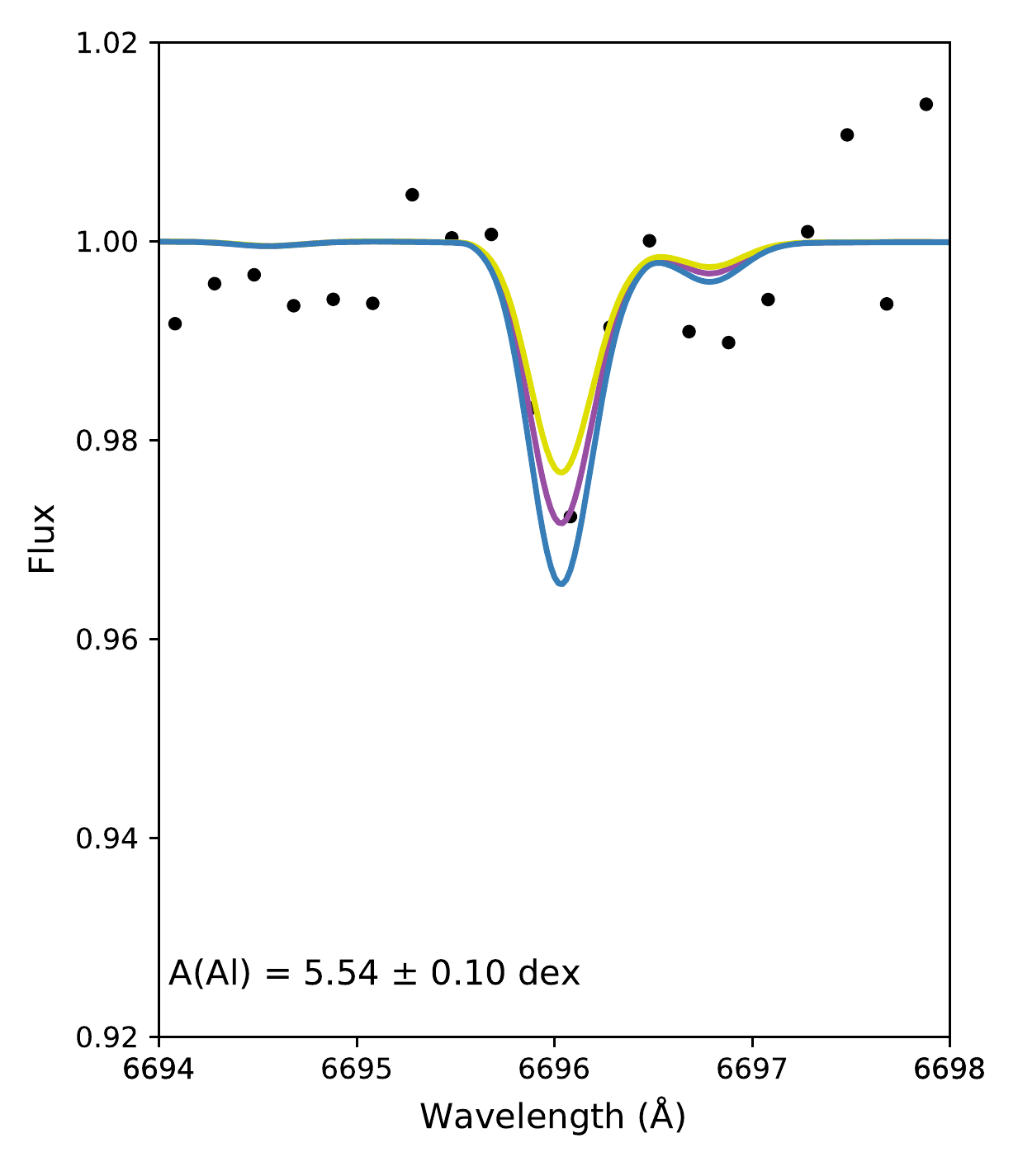}
         \label{fig:al_line}
     \end{subfigure}
     \begin{subfigure}[b]{0.45\textwidth}
         \centering
         \includegraphics[width=\textwidth]{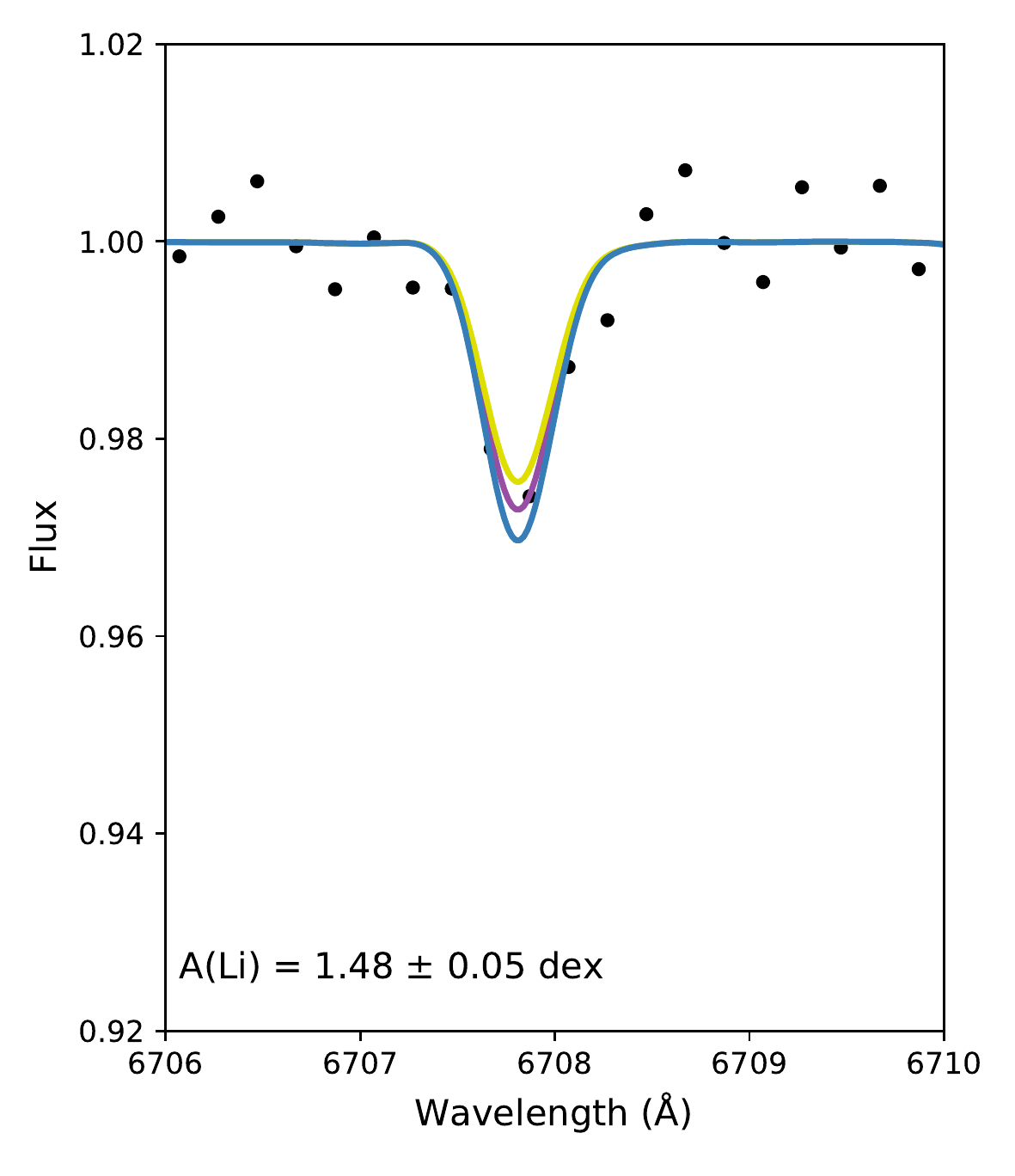}
         \label{fig:li_line}
     \end{subfigure}
     \caption{On the left (right): Instance of aluminium (lithium) line synthesis for a star member of NGC 6752. Black dots show the observed spectrum, yellow, purple, and blue lines are the synthetic spectra: for aluminium 5.44 dex, 5.54 dex, 5.64 dex, and for lithium 1.38 dex, 1.48 dex, and 1.58 dex, respectively. The best fit (purple line) corresponds to A(Al) $= 5.54$ dex and A(Li) $= 1.48$ dex.}
  \label{fig:fitting_line}
\end{figure*}

\subsection{Measurement of Li and Al}

\citet{Cayrel1988} (his equation 7) describes how to derive the uncertainty associated with EW measurements based on spectral resolution and signal-to-noise ratio. As \citet{Mucciarelli2011} adopted it, we assumed three times this uncertainty as the minimum measurable EW. In the neighbourhood of the wavelength corresponding to the relevant transitions (Li and Al), this value range from 8 m\r{A} to 30 m\r{A}, and from 3 m\r{A} to 5 m\r{A}, for GIRAFFE and UVES spectra, respectively. We established that Li and Al are measurable in 153 and 34 stars, respectively, based on these cutoffs. In contrast, no meaningful measurement can be derived for the remaining stars, but only upper limits.

Li and Al abundances were measured via spectral synthesis using the \texttt{synth} driver from \texttt{MOOG}. The line list adopted is the one used by \citet{Dorazi2015b}.
We corrected our Li measurements to NLTE abundances using corrections obtained from the \texttt{INSPECT} database version 1.0\footnote{\url{http://inspect-stars.com/}}, which are based on the ones provided by \citet{Lind2009}. These corrections range from $\sim-$0.02 dex to $\sim-$0.06 dex for unevolved and evolved stars, respectively. The Al lines used are also affected by NLTE effects; hence we applied the corrections given by \citet{Nordlander2017}. While the Al corrections are quite small for evolved stars (mean correction of 0.02 dex), the unevolved stars are more affected by this phenomenon (mean correction of 0.09 dex).

Figure \ref{fig:fitting_line} shows examples of spectral synthesis fitting to the Al (6696 \r{A}) and Li doublet at (6708 \r{A}). The Al line at 6698 \r{A} is expected to be $\sim$2 times weaker than the one at 6696 \r{A}, which is already weak in our spectra; hence our results are based on the measurements done in the latter. Given the weakness of the Al line, the fitting of those lines has large uncertainties in many of the objects in our sample. However, the strictly uniform procedure followed for all the analysed spectra minimises the random errors due to fitting and atmospheric parameters uncertainties.

The purple lines indicate the best fit, which corresponds to a A(Al) = 5.54 dex\footnote{A($X$) $= \log$(N$_{X}$/N$_{H}$)+12, where N$_{X}$ is the number density of the relevant species.} and A(Li) = 1.48 dex. Two more synthetic models are shown in yellow and blue, with an aluminium and lithium abundance difference with respect to the best one of 0.10 dex and 0.05 dex, respectively.

\subsection{Observational uncertainties}

The uncertainty associated with the measurements combines of the uncertainties of the best-fit determination and those associated with the uncertainties in the adopted atmospheric parameters. To determine them, we followed the approach described by \citep{dorazi2014}.
We changed the stellar parameters one at a time, keeping fixed the remaining ones and reporting the corresponding variation in both Li and Al. 
We selected two stars to represent the sample: \#3081 for TO/SGB and \#2245 for RGB ones to estimate the sensitivity of the abundances measured to the change of stellar parameters. The variations assumed in stellar parameters are: $\Delta$T$_{\mathrm{eff}} =  100$ K, $\Delta$$\log g=  0.2$ dex, $\Delta$v$_{m}$$= 0.1$ km s$^{-1}$, and $\Delta$[Fe/H]$ =  0.1$ dex. 
This sensitivity matrix can be found in the table \ref{tab:sensitivity_matrix}. The errors listed in Table \ref{tab:summary_1} in columns 11 and 13 take into account the uncertainty due to the error in both the atmospheric parameters and in the fitting.

\begin{table*}
   \centering
   \caption{Elements sensitivity to the change in stellar parameters on representatives of TO/SGB star (\#3081) and RGB star (\#2245).}
   \label{tab:sensitivity_matrix}
   \begin{tabular}{cccccc}
       \hline
       \hline
       ID & Element & T$_{\mathrm{eff}}$ (+100) & $\log g$ (+0.2) & v$_{t}$ (+0.1) & {[}Fe/H{]} (+0.1) \\
       \hline
\multirow{2}{2em}{\rotatebox{0}{\#3081}}  
       & A(Li)   & 0.08            & 0.01           & 0.01          & 0.01             \\
       & A(Al)   & 0.02            & 0.01           & 0.01          & 0.00             \\
       \hline
\multirow{2}{2em}{\rotatebox{0}{\#2245}}  
       & A(Li)   & 0.10            & 0.00            & 0.01          & 0.01             \\
       & A(Al)   & 0.05            & 0.01           & 0.01          & 0.01 \\
       \hline
\end{tabular}
\end{table*}

Table \ref{tab:summary_1} list the stellar parameters, the abundances measured, and their respective errors \footnote{The full table is only available in electronic form.}, while Table \ref{tab:summary_2} shows a summary with the number of analysed targets in this work.

\subsection{Comparison with the literature}

\citetalias{Gruyters2014} reported (22) 99 measurements and (48) 25 upper limits for (Al) Li out of the 126 stars that spectra in the wavelength range covering Al and Li. Here we compare the measured Li and Al for these stars.

\begin{figure*}
	\includegraphics[width=2\columnwidth]{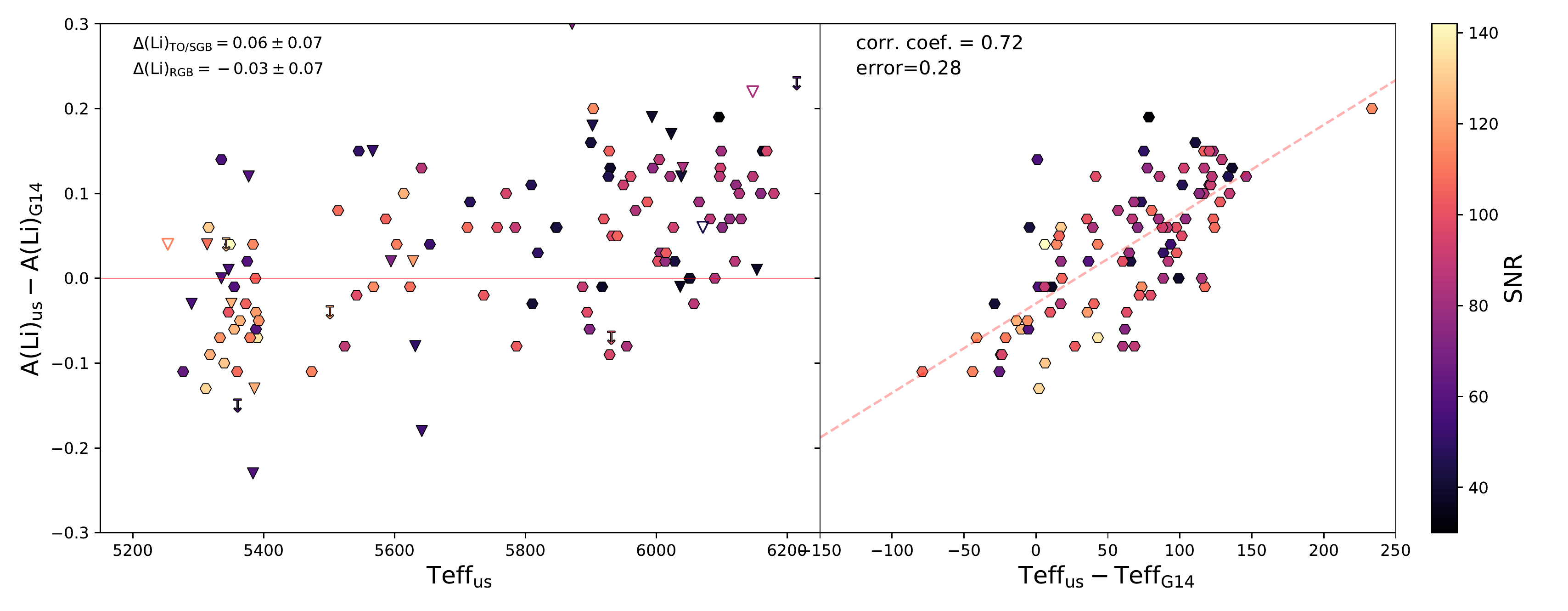}
    \caption{Difference between our Li abundances and the ones reported by \citetalias{Gruyters2014} as a function of $\mathrm{T_{eff}}$ (left), and as a function of the difference between our $\mathrm{T_{eff}}$ and the ones reported by \citetalias{Gruyters2014} (right). Filled hexagons and down-pointing symbols represent measurements and upper limits, respectively. Empty triangles, filled triangles, arrows show that we measured, and they did not; they and we reported upper limits, we report an upper limit, and they an actual measurement. Colours vary with the SNR.}
    \label{fig:diff_li}
\end{figure*}

\begin{figure*}
	\includegraphics[width=2\columnwidth]{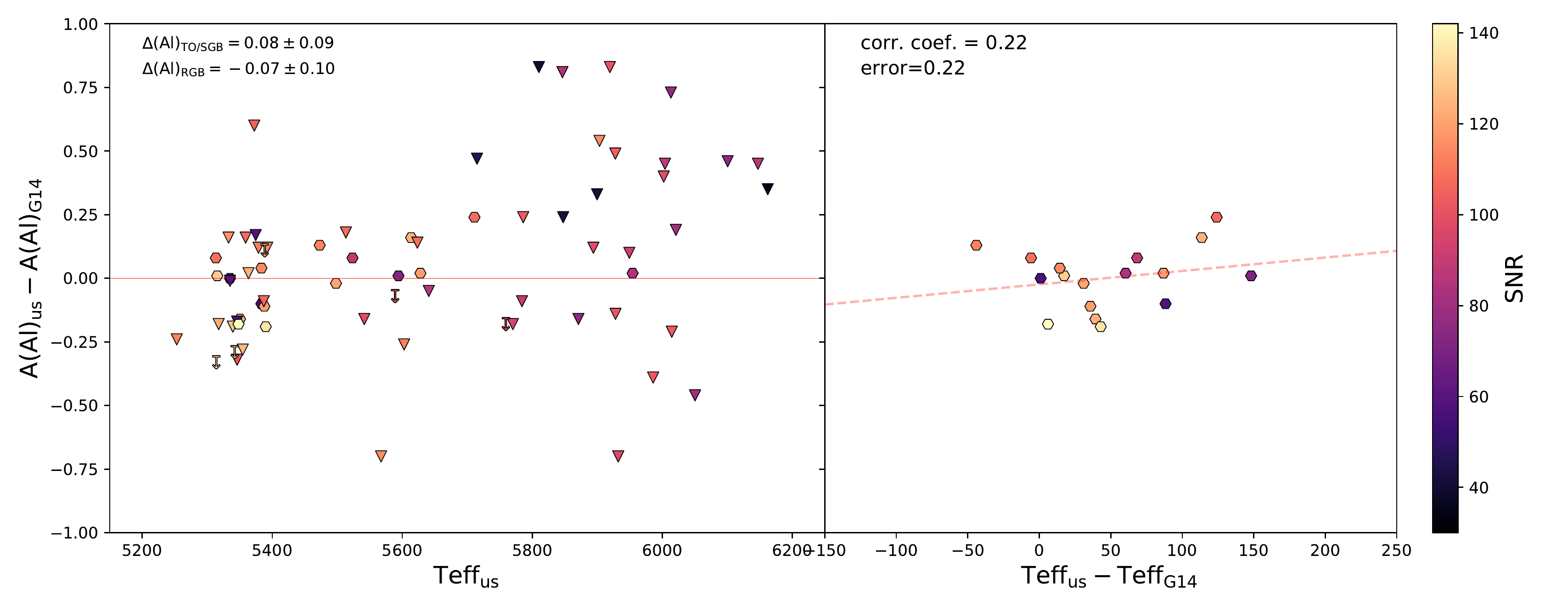}
    \caption{Difference between our Al abundances and the ones reported by \citetalias{Gruyters2014} as a function of $\mathrm{T_{eff}}$ (left), and as a function of the difference between our $\mathrm{T_{eff}}$ and the ones reported by \citetalias{Gruyters2014} (right). Filled hexagons and down-pointing symbols represent measurements and upper limits, respectively. Empty triangles, filled triangles, arrows show that we measured, and they did not; they and we reported upper limits, we report an upper limit, and they an actual measurement. Colours vary with the SNR.}
    \label{fig:diff_al}
\end{figure*}

Figures \ref{fig:diff_li} and \ref{fig:diff_al} compare our measurements for Li and Al and the ones from \citetalias{Gruyters2014} as function of T$_{\mathrm{eff}}$. The best linear fit (red dashed line) is shown on the right panels with their respective correlation coefficient.
In both cases, there is an indication of a trend. The difference between the Li measurements presented here and those in G14 are strongly correlated with the difference in $\mathrm{T_{eff}}$, as shown in the right panel of Fig \ref{fig:diff_li}. It is noteworthy that the most discrepant points in this plot correspond to stars for which the spectra had poor SNR.

We note that unlike for Li, the trend in Al, in particular for unevolved stars, is based on very few stars. Then, it should be taken with great caution. Still, the measurements of Al abundances are similar in both studies.
These comparisons indicate that the differences in Li and Al abundances between the present paper and G14 are consistent with the differences in adopted \teff.

\section{Results and Discussion}
\label{Sec:Results}

\subsection{Al distribution}

\begin{figure}
	\includegraphics[width=\columnwidth]{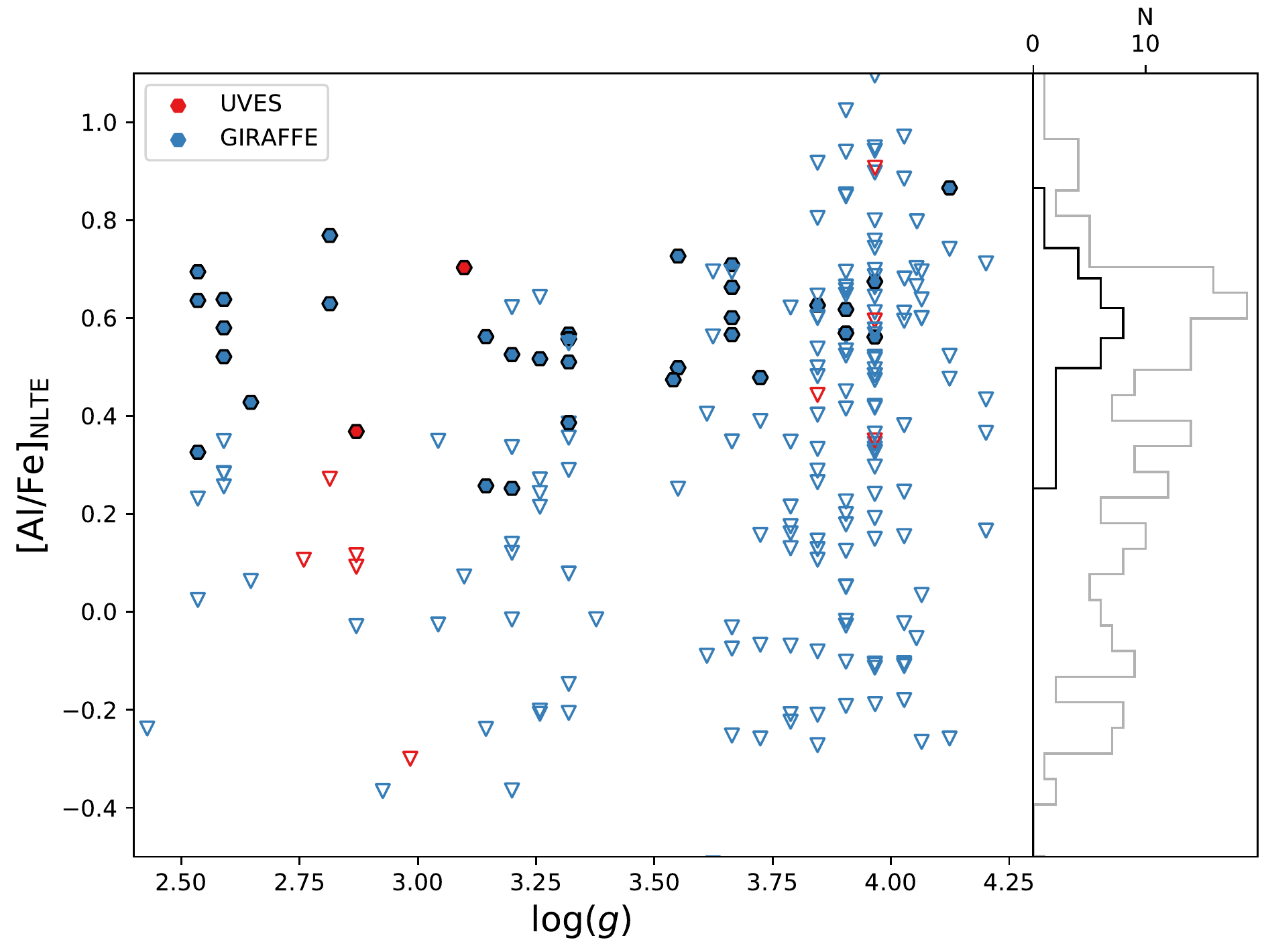}
    \caption{[Al/Fe] as a function of $\log g$. Blue and red symbols represent GIRAFFE and UVES spectra. Empty down-pointing triangles represent upper limits in Al. The black and grey histograms on the right show the counts of actual measurement (only filled hexagons) and the whole sample, respectively.}
    \label{fig:al_teff}
\end{figure}

\begin{figure}
	\includegraphics[width=\columnwidth]{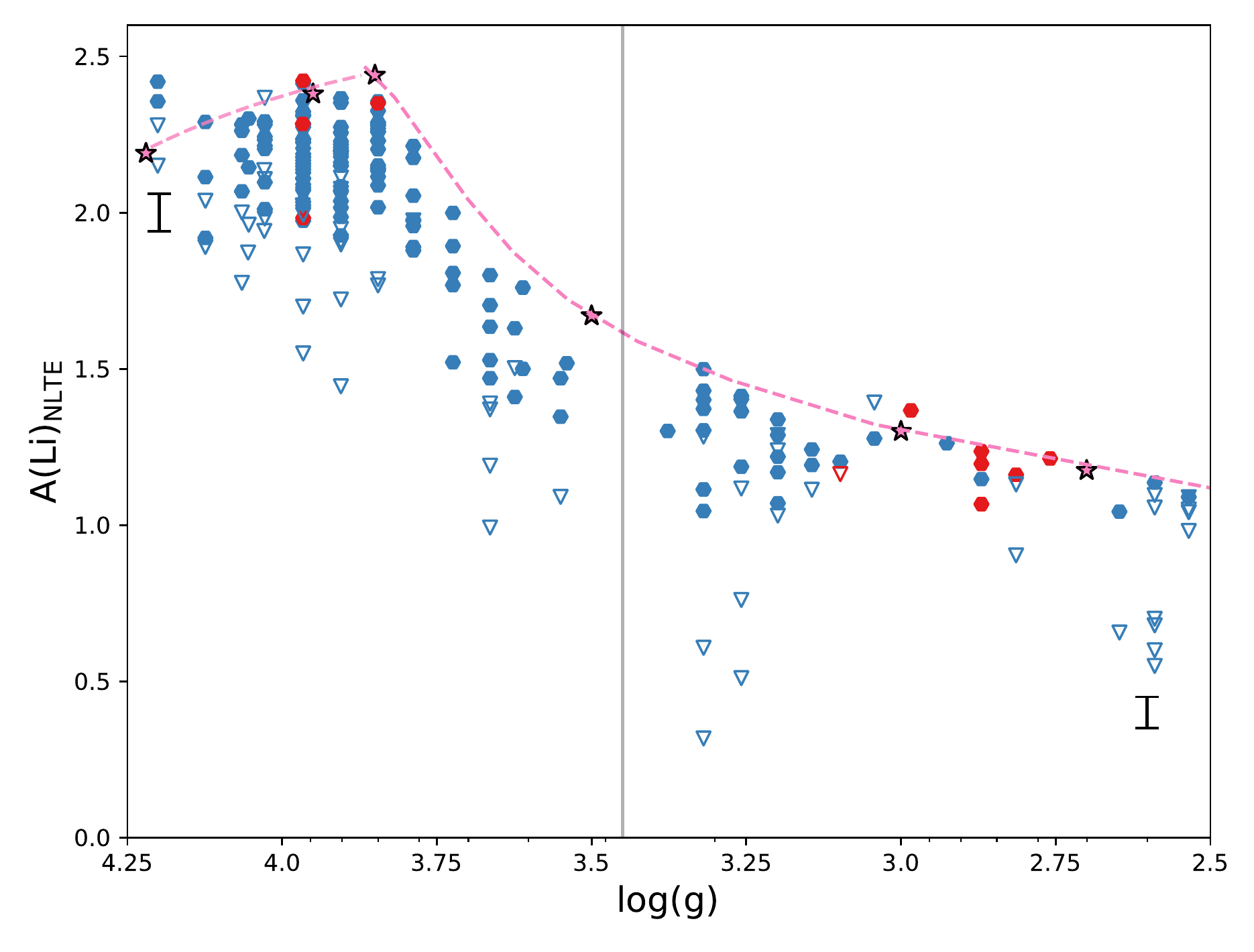}
    \caption{Li$_{NLTE}$ abundance as a function of the $\log g$. Red and blue symbols show UVES and GIRAFFE spectra, respectively. Closed hexagons are measurements, while open triangles represent upper limits in Li. The pink stars and the dashed line indicate the expected evolution of Li. The vertical solid line shows the $\log g$ at the bottom of the RGB.}
    \label{fig:li_teff}
\end{figure}

\begin{figure}
	\includegraphics[width=\columnwidth]{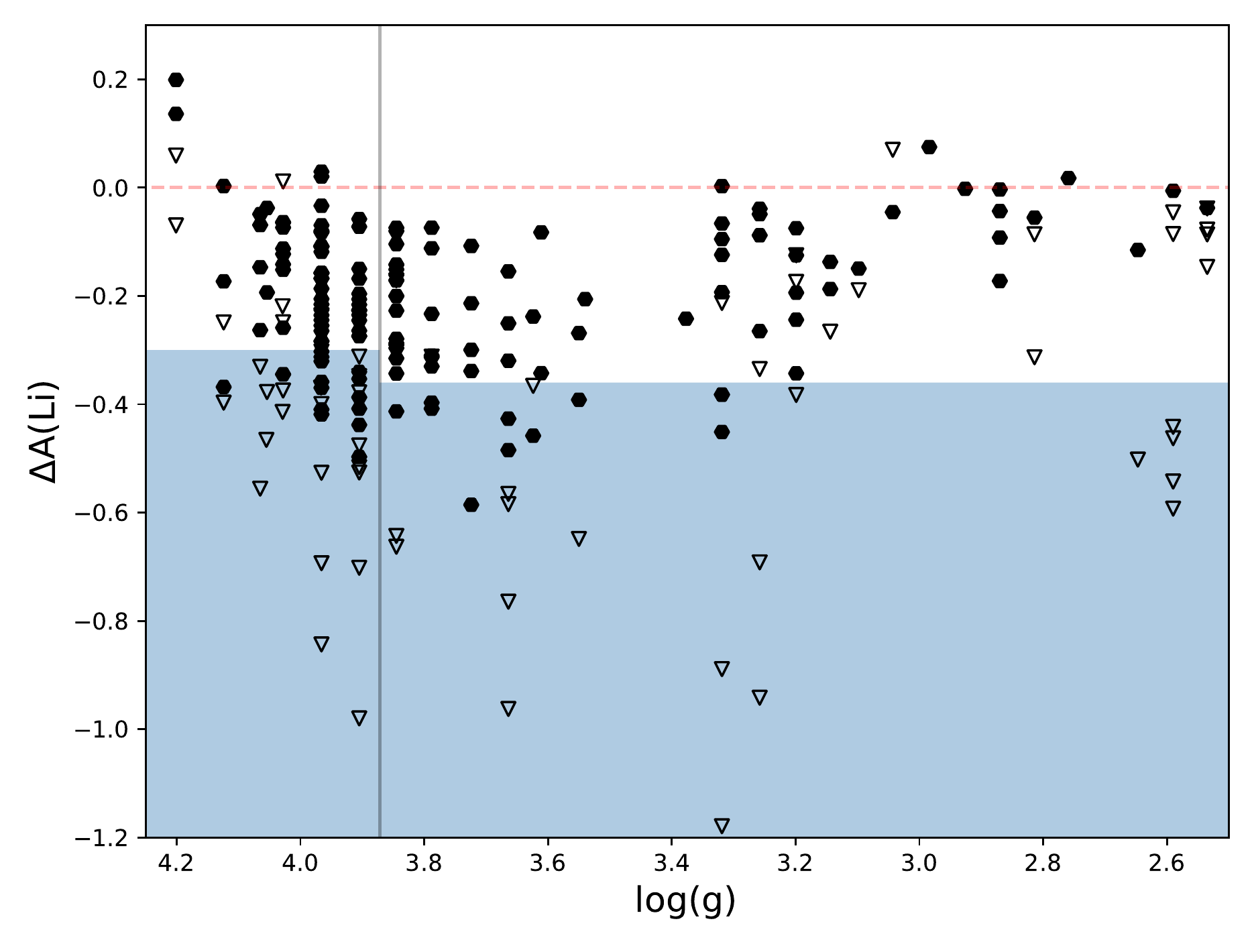}
    \caption{$\Delta$(Li)--$\log g$ plane. The blue area shows the values larger than >6$\sigma$ from the fit. Symbols follow the same description as Fig. \ref{fig:li_teff} for Li.}
    \label{fig:Deltali_teff}
\end{figure}

\begin{figure}
     \centering
     \begin{subfigure}[b]{\columnwidth}
         \centering
         \includegraphics[width=\textwidth]{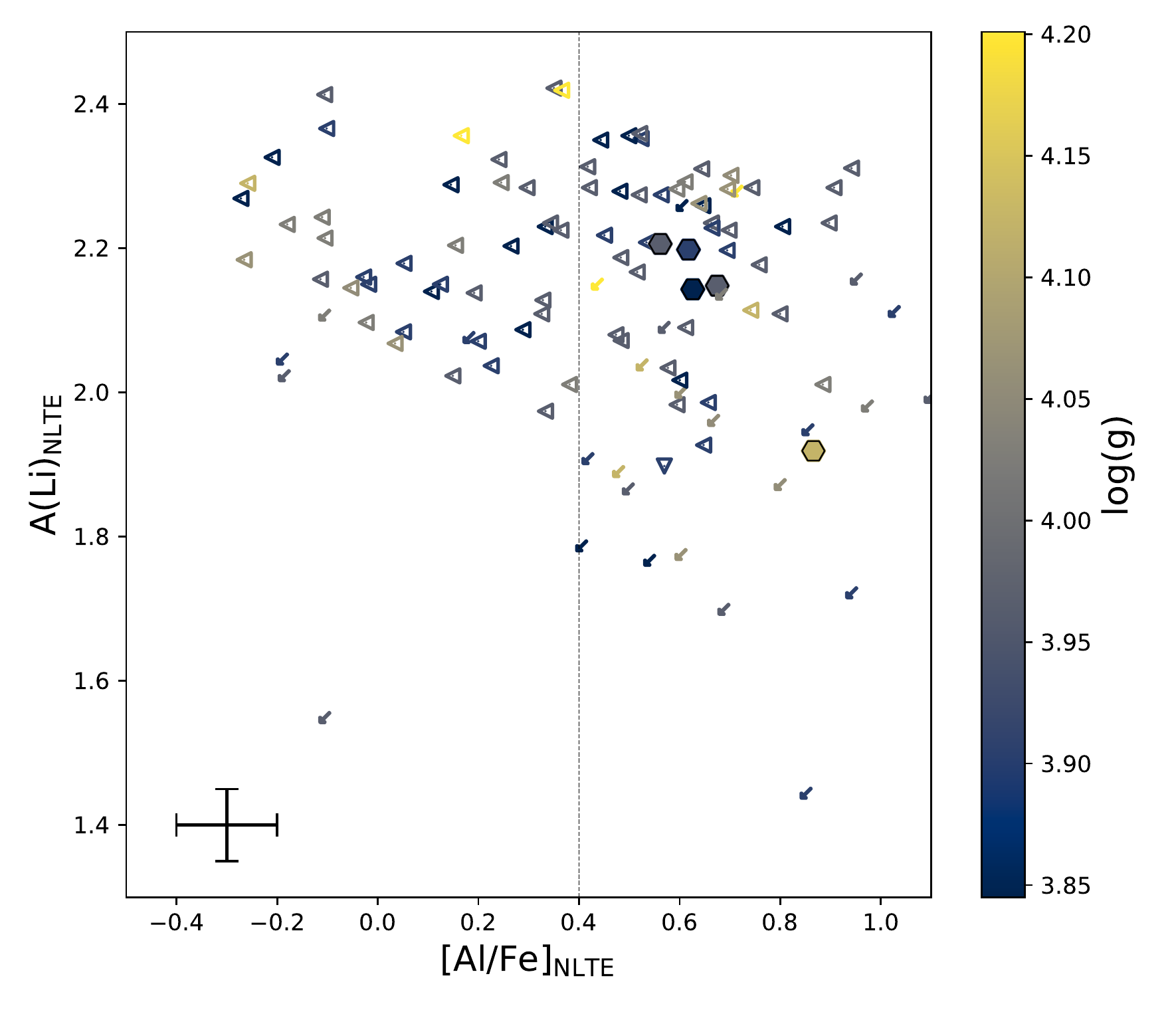}
         \label{fig:li_al1}
     \end{subfigure}
     \begin{subfigure}[b]{\columnwidth}
         \centering
         \includegraphics[width=\textwidth]{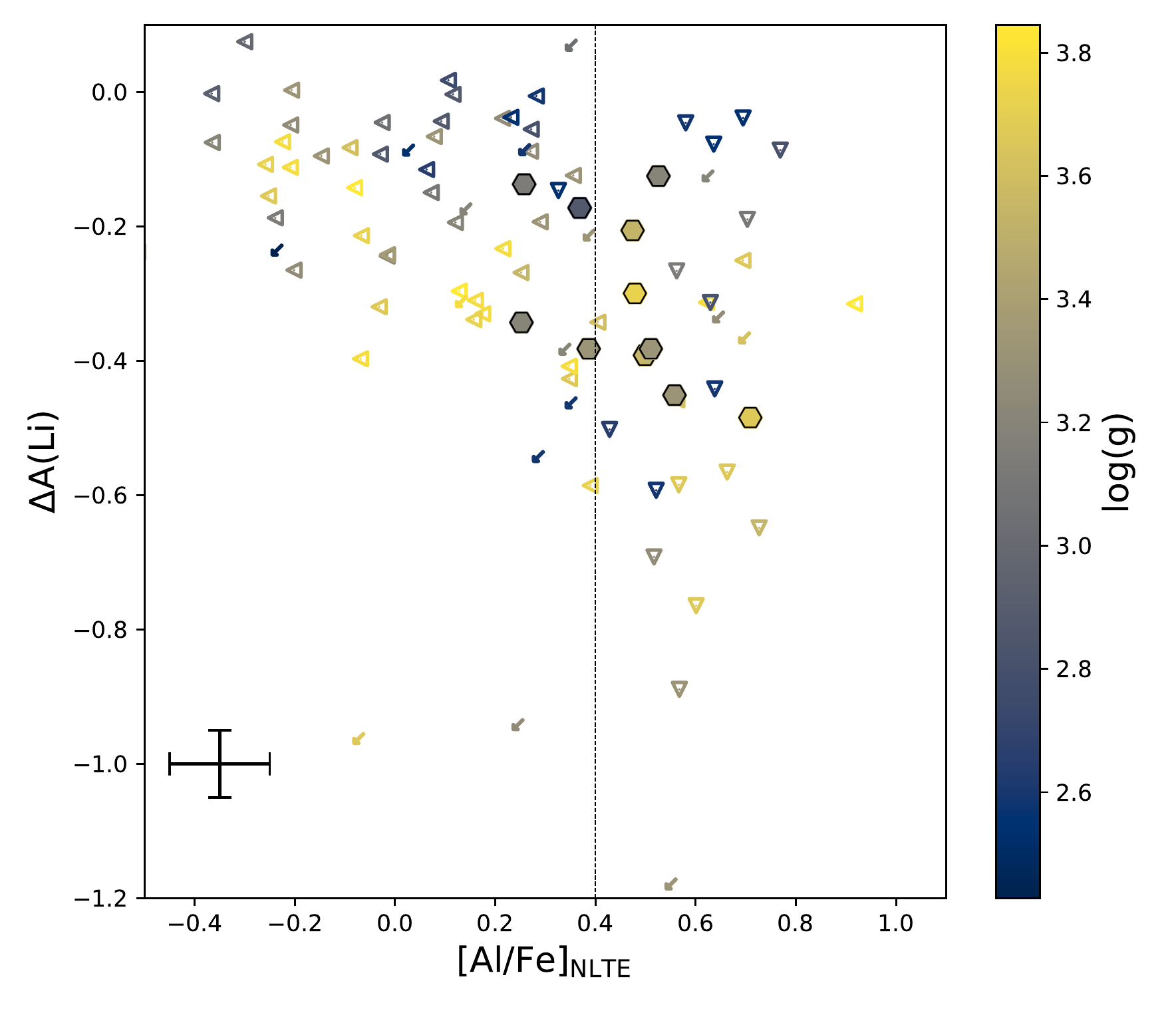}
         \label{fig:li_al2}
     \end{subfigure}
     \caption{Relation between aluminium and lithium abundances. Upper and lower panels show the results for stars with higher and lower $\log g$ than 3.87, respectively. Symbols are coloured based on the $\log g$ scale. The vertical dashed line at [Al/Fe]$=0.4$ dex is the threshold value used by \citep{Carretta2012} to separate the first and second generation in NGC 6752. Left-pointing triangles show upper limits in Al. Diagonal arrows represent upper limits in Li and Al. Finally, down-pointing triangles represent upper limits in Li.}
  \label{fig:li_al}
\end{figure}

Figure \ref{fig:al_teff} shows the distribution of Al measurements and upper limits as a function of $\log g$. Detection was possible only in Al-rich stars, while most stars have an Al content too low to produce a detectable line at the temperature and signal-to-noise ratio of the star.
Our results show a dispersion of Al content in our sample, consistently with previously published studies \citep{Meszaros2020,Carretta2012,Carretta2009}. The spread is found at all evolutionary stages.

Considering upper limits and detections, we can estimate that the Al spread is at least 1.2\,dex. It is worth noticing that there is no evidence of an offset in the Al content between TO/SGB and RGB stars, which argues in favour of our approach in the derivation of a uniform temperature scale for the whole sample. 
The Al spread is lower than the spread reported by \citealt{Carretta2012} ($\sim$1.75 dex), but it is in good agreement with \citealt{Carretta2009}, who reported an Al spread of $\sim$1.11 dex. This could be due to the fact that the former is based on a different set of lines (Al doublet at  8772–8773\AA), while the second uses the same set of Al lines as us.

The distribution of the Al abundance is of interest: in the context of the multiple-generation scenario for the MSP, \citep[see e.g.][]{Carretta2013}, a clumpy distribution would argue in favour of different episodes of star formation, while a smooth one would be consistent with continuous star formation.
In our sample, we do not find any evidence for clumpy distribution, as can be seen from Fig \ref{fig:al_teff}.  Note that this is not inconsistent with previous findings of a clumpy distribution (see e.g., \citealt{Carretta2012}) as there are only a small number of actual Al detections (as opposed to upper limits) in this sample.

\subsection{Li evolution and model predictions}

Figure \ref{fig:li_teff} shows the distribution of A(Li)$_{NLTE}$ as a function of the $\log g$.
The overall Li abundance shows a continuous decrease from a plateau at A(Li)$_{NLTE}$=2.33$\pm$0.06 dex, where TO stars show the highest Li abundance. 
We estimated this value by averaging the measurements for the stars in the upper quartile of the distribution of Li measurements, with $\log g$ > 3.87. 
This choice was determined by the need to avoid the influence of Li depleted stars. This plateau is in agreement with previous studies (\citealt{Mucciarelli2011}, \citetalias{Gruyters2014}), showing a significant difference from the lithium abundance expected from the Big Bang nucleosynthesis. This discrepancy goes beyond the scope of the present paper, but we refer the interested readers to the relevant literature on the topic, e.g., \citet{Fu_2015,Matteucci_2021}. 

The observed drop of the lithium abundance with decreasing $\log g$ is expected based on stellar evolution. After stars leave the main sequence, stars go through their sub-giant branch phase and climb up the RGB. Then they experience dredge-up episodes and mixing process, which bring to the stellar surface material processed in layers where the temperature is high enough to burn Li, decreasing the  Li content, until it is essentially completely destroyed after RGB-bump. 
\citet{Gratton2000} showed the depletion of Li observed in metal-poor field stars as they evolve, finding however negligible dispersion among stars in the same evolutionary stage. 

We used six different evolutionary tracks from \citet{Pietriferni2021} to derive the theoretical predictions for Li abundances and evolutionary depletion. The tracks stellar masses are 0.70\msun, 0.75\msun, 0.80\msun, 0.85\msun, 0.90\msun, and 1.0\msun, which correspond to stars between the TO and the bottom RGB for a cluster of 12 Gyr and a [Fe/H]=-1.55 dex. The expected Li abundance for a GC of this age, for each track at a given $\log g$ is shown with pink stars in Fig.\ref{fig:li_teff}. The dashed line interpolates these points by fitting a first and sixth-order function for unevolved and evolved Li prediction, respectively. Note that the fitting of the evolutionary Li is not intended to be rigorous, but to follow the Li variations at different evolutionary stages. Although the overall Li predicted is overestimated, the upper envelope of the Li observed of FG stars tends to follow the prediction closely, indicating a good agreement with the models. It is worth noticing that the Li predictions are pretty sensitive to the models' mixing processes. The good match between predictions and observations supports the treatment of mixing processes in the models.
As shown in Fig. \ref{fig:li_teff}, in addition to the evolutionary Li decrease, at each $\log g$, there is a considerable range of Li, suggesting that there is an additional factor involved. This is peculiar to GCs, and it is a behaviour thought to be related to the MSP phenomenon present in GCs.

In order to better probe the variations of Li not due to evolution, we have to account for the latter. We have thus defined the quantity $\Delta$A(Li), the difference between the Li measured and its corresponding Li from the fitting line (see Fig. \ref{fig:li_teff}) at every $\log g$.

Figure \ref{fig:Deltali_teff} shows the $\Delta$A(Li)--$\log g$ plane. 
If the only factor determining Li depletion were evolution, then the expectation is that stars would follow the red dashed line of Fig.\ref{fig:Deltali_teff}. As mentioned earlier, \citet{Gratton2000} showed that field stars behave in such a fashion. A large fraction of stars in our sample is considerably depleted in Li for their evolutionary status, a characteristic that is peculiar to GCs.

It is interesting to look at a the fraction of Li-poor stars: \citet{Gratton2019} found a strong correlation between the fraction of Li-poor population of a GC with its extreme fraction of stars as determined by \citet{Carretta2009b}, which is 0.4 for this cluster. We adopt as a working definition of Li-poor stars that of being 6$\sigma$ below the theoretical prediction.
In Fig.\ref{fig:li_teff} the blue shading shows the area corresponding to the >6$\sigma$ from the fit. Note that it is split in unevolved and evolved stars, adopting $\log g$=3.87 (grey line) as separation, as it is the point where the overall Li abundance starts being affected by evolutionary depletion  (see Fig. \ref{fig:li_teff}). 
Based on this definition, the fraction of Li-poor stars is 0.43$\pm$0.05.  This result is in good agreement with the expected fraction of Li-poor stars shown in Fig. 18 of \citet{Gratton2019}, and in better agreement than that reported by \citet[]{Shen2010}, 0.30$\pm$0.05. However, they analysed only unevolved stars, in which, due to their temperatures, Li is measurable only when relatively high, leading to a selection effect. 

It is clear here that there is an intrinsic spread at every evolutionary stage, which we expect to be related to the MSP phenomenon. Moreover, the spread seems roughly the same even beyond the TO, suggesting that the stars had different Li content from their birth, and the spread is maintained through the depletion due to stellar evolution.

\subsection{Li and Al}

It is interesting now to examine the relation between Li and Al in the cluster's stars.
Figure \ref{fig:li_al} shows the Li abundance as a function of [Al/Fe]. The upper panel shows our results for stars where the evolutionary effect on Li has not taken place yet, while the bottom panel is for evolved stars.

\citet{Carretta2012} considered [Al/Fe] = 0.40 dex as the threshold between two populations, with the first- and second-generation stars being the ones with smaller and larger aluminium abundance than the threshold value, respectively. However, the threshold is based on a distribution derived with a different set of aluminium lines (the Al doublet 8772–8773\,\AA), and thus in principle, not fully consistent with our Al abundances, which are based on the  6696 and 6698\,\AA lines. In fact, \citet{Carretta2012} obtained a larger Al abundance spread ($\sim$1.5 dex) than the present ones or that of \citet[]{Carretta2009} (also based on the 6696 and 6698\,\AA lines).
In order to investigate potential offsets in the Al distributions derived using the two sets of lines mentioned above, we examined the Al abundances in 3 stars in common between \citet{Carretta2009} and \citet{Carretta2012}. Note that we have no star in common with either of these studies, so a direct comparison was not possible.
We find that [Al/Fe]=0.4 dex as derived from the 8772 and 8773\,\AA lines corresponds to 0.34$\pm$0.07 dex derived from the lines used in this work. Given that only three stars were available for this exercise and that the rescaled Al threshold value is consistent with the original one within the error, we adopt the 0.4 \citet{Carretta2012} value as the threshold between the first and second-generation stars in the rest of the discussion.

In the upper panel of Fig. \ref{fig:li_al}, the measurements of Al and the upper limits on the left-hand side show that there is a considerable Al spread even among stars with high Li (A(Li)$\sim$ 2.25). In other words, we found Li-rich stars both among first- (Al-poor) and second- (Al-rich) generation stars.  This is not an evolutionary effect, nor is it expected from the pure pollution scenario from hot H-burning processed (Na, Al, N rich and C, Mg and Li poor) material. 

In the lower panel, we probe the same relation for evolved stars. As this sample includes stars that have experienced various amounts of evolutionary Li depletion, to single out the effect of MSP, we considered the previously defined term, $\Delta$A(Li) instead of the A(Li). After the evolutionary effect has been removed, besides Li-rich, Al-poor stars, and Li-poor, Al-rich stars, there is also a number of Al-rich, SG stars, with the same Li content of their FG, Al-poor counterparts.

\begin{figure}
	\includegraphics[width=\columnwidth]{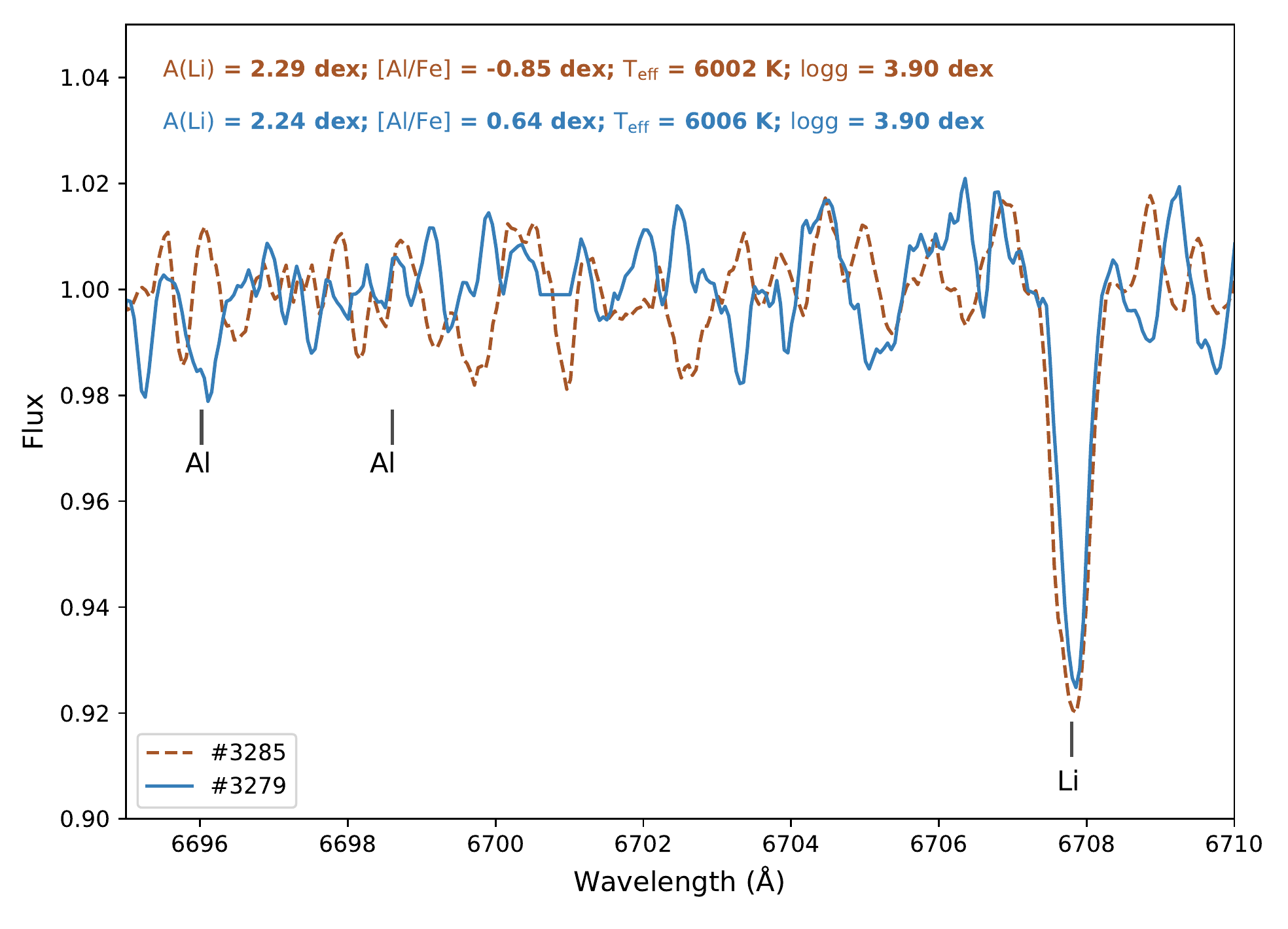}
    \caption{Comparison between star \#3285 (dashed brown line) and star \#3279 (solid blue line). Both have similar stellar parameters with similar Li abundance. While could only place an Al upper limit for the star \#3285, the star \#3279 shows a higher Al abundance of [Al/Fe]=0.64 dex. We recall to the reader that our result are based on just Al line at 6696\r{A}.} 
    \label{fig:comparison_spectra}
\end{figure}

In Fig. \ref{fig:comparison_spectra}, we show a comparison between stars with similar stellar parameters and the same Li abundance (within the errors). Their quite different Al abundances show that they belong to different populations.

As was discussed in the introduction, the simplistic expectation is that the content of Li and p-capture elements should be anticorrelated. The stars could be divided into: i) a Li-rich and Al-poor population (so-called first-generation stars) and ii) a Li-poor and Al-rich population (so-called second-generation stars). These stars are, in fact, found in both unevolved and evolved stars. However, in addition, we find a third group, an unexpected portion of stars that have both high aluminium and lithium. 

As it is predicted for the models, in a GC of this metallicity, the initial Li abundance decreases by a factor of $\sim$20 after the first dredge-up (e.g.,\citealt[]{Mucciarelli_Salaris2012}). This drop is consistent with the drop that we found in the evolved FG stars with respect to unevolved FG stars. On the other hand, although Li upper limits dominate the second group, the results are consistent with stars created from a material with low lithium content or even from Li-free material. The chemistry of this group is consistent with having formed from material enriched by p-capture rich, Li-poor material. Finally, the stars in the third group suggest the presence of some source of Li production, which enriches a fraction of the SG stars. It is interesting to note that the Li abundance never exceeds that of the plateau. The Li production seems to at most compensate the Li destruction, and it does not reach a kind of forbidden Li zone above such abundance value. It is worth noticing that all these three groups are visible in both panels.

Among the candidate polluters for GCs, the only nucleosynthetic site known to produce Li are intermediate-mass AGB stars \citep[][]{Gratton2019} and our findings suggest that these stars must have contributed to the pollution of the observed stars in the cluster under discussion. The same conclusion was reached for other clusters (see e.g., \citealt{dorazi2014} for NGC~6218 and \citealt{Dorazi2015} for NGC~362). Moreover, the fact that we are finding different Li abundances among stars with similar Al abundance in the same evolutionary stage hints at the possibility of the Li production in just a subset of the polluters. Probing the content of other chemical species could provide further insight on this issue. This will be the focus of an upcoming paper.

According to models, AGB stars can produce the p-capture elements involved in the anti-correlations and Li \citep[]{Ventura2008,Ventura_2009}, which is produced through the Cameron-Fowler mechanism \citep{Cameron1971}, in non-negligible amounts. Therefore, in principle, these polluters can explain the observed patterns not only for what concerns C, N, O, Na, Mg, and Al, but also Li elements. 

However, it must be kept in mind that the details of the Li production are quite uncertain, so the quantitative predictions should be taken with some caution.
The production of Li varies strongly with the mass of the AGB star, as shown in \citet{Dantona_2019} who modelled the Li yields in different AGB masses. They used NGC~2808 as a GC prototype to explain the MSP phenomenon in GCs. They claimed that the chemical abundances of non-extreme SG stars could be explained by the diluted pollution of AGB stars, of different masses, with the pristine gas.

We note that Li-rich giants (with A(Li) larger what expected for their evolutionary stage by $\sim$0.5-3.0 dex) are a known and rare phenomenon found both in the field and stellar clusters (e.g., \citealt{Sanna2020,Mucciarelli2021}). The nature of these rare objects is still under debate. They are not expected to be related to the MSP phenomenon, but to an evolutionary effect. We refer the reader to the mentioned papers for a detailed discussion. Given the rarity of the phenomenon, it is unlikely that classical Li-rich giants account for a non-negligible fraction of the high Li, SG evolved stars we are observing. In fact, there is no evidence of an excess of Li-rich stars among evolved stars with respect to unevolved stars. However, it is important to keep in mind that the minimum measurable Li (or the minimum value of upper limits) in dwarves is higher than in evolved stars.

It has also been reported (e.g., \citealt{Mori2021,Magrini2021}) that red clump stars undergo a Li enhanced phase just after the upper RGB, and in principle, they could also contribute to the Li production in the cluster. However, these Li enhancements are not expected at the metallicity of GC and at the evolutionary stages of our sample. Besides, even at higher metallicities, they are quite modest ($\sim$0.6 dex; \citealt{Zhang2021}), not enough to explain the Li abundances found in our sample.

Finally, some studies have argued that the observational evidence in several clusters can not be reconciled with a single class of polluter (e.g., \citealt{Carretta_2018}). The present data do not allow speculating further on this issue. However, they provide a strong indication that AGB stars must have contributed to the pollution of the SG stars also in NGC~6752.

\section{Summary and Conclusions}
\label{Sec:Conclusion}

We analysed 217 stars in metal-poor GC NGC~6752. We performed a spectrum-synthesis on FLAMES spectra in GIRAFFE and UVES mode and derived homogeneous abundances of Li and Al in stars from the TO to the RGB. These species were studied to shed light on the polluter(s) responsible for the chemical patterns observed in SG stars in the cluster. 

Our sample shares 126 stars with \citetalias{Gruyters2014}, and we are in good agreement with their findings, with differences consistent with the slightly different temperature scale adopted.

We reported 153 Li measurements and 64 upper limits. Additionally, we reported 183 Al upper limits. The aluminium measurements were possible only in a small fraction of the sample (34 stars) where the Al line was strong enough to be detected. Although the sample is quite small, the Al distribution does not seem to be bimodal.

The Li abundances have an overall decrease from the TO to the RGB. The drop is expected because of stellar evolution.
To disentangle the last effect from the MSP phenomenon in evolved stars, we defined $\Delta$(Li). In this way, we could detect the presence of a Li-Al anti-correlation, which was also found in unevolved stars.

We detect Li-rich stars among both first- and second-generation stars, with generations defined on the bases of Al as in \citep{Carretta2012}.

The detection of Li-rich and Li-poor stars among SG (Al-rich) stars indicates the need for Li production, which is known to happen in AGB stars. While our data imply that AGB stars have contributed to the pollution of the SG stars in NGC~6752, they provide no insight into whether or not other sources of self-pollution (e.g., FRMS, supermassive stars, or massive binaries) were involved in the chemical evolution of the cluster.
Future work on a more extensive set of elements, including heavy elements, could, in principle, help disentangle this issue.

\section*{Acknowledgements}
We thank our referee Achim Weiss for his helpful comments and recommendations.

J.S-U and his work were supported by the National Agency for Research and Development (ANID)/Programa de Becas de Doctorado en el extranjero/ DOCTORADO BECAS CHILE/\texttt{2019-72200126}.

M.J.R. and her work were supported by the National Agency for Research and Development (ANID)/Programa de Becas de Doctorado en el extranjero/ DOCTORADO BECAS CHILE/\texttt{2018-72190617}.

This work was partially funded by the PRIN INAF 2019 grant ObFu 1.05.01.85.14 ('Building up the halo: chemo-dynamical tagging in the age of large surveys', PI. S. Lucatello)

We thank V. D'orazi for providing the line list and the useful discussion, P. Gruyters and A. Korn for the data, F. Grundahl for sharing with us the photometry of the cluster, S. Cassisi for providing us the evolutionary tracks, and T. Nordlander to share his grids of Al corrections.

\section*{Data Availability}
The spectra used in the article are available on the ESO Science Archive Facility (\url{http://http://archive.eso.org/wdb/wdb/adp/phase3\_spectral/form}) under the ID programs 077.D.0246(A), 079.D-0645(A), 081.D-0253(A), and 083.B-0083(A). The data provided by \citet[]{Gruyters2014} can be found from the same site under the ID programs 079.D-065(A) and 081.D-0253(A). Str\"{o}mgren photometry is from \citep{Grundahl1999} and it was kindly provided by the author. Last, Gaia eDR3 data are available on \url{https://gea.esac.esa.int/archive/}.



\bibliographystyle{mnras}
\bibliography{mnras_template} % if your bibtex file is called example.bib

%%%%%%%%%%%%%%%%%%%%%%%%%%%%%%%%%%%%%%%%%%%%%%%%%%

%%%%%%%%%%%%%%%%% APPENDICES %%%%%%%%%%%%%%%%%%%%%

%\appendix


%%%%%%%%%%%%%%%%%%%%%%%%%%%%%%%%%%%%%%%%%%%%%%%%%%

% Don't change these lines
\bsp	% typesetting comment
\label{lastpage}
\end{document}